# Nanostructures and Defects in Non-equilibrium Synthesized Filled Skutterudite $CeFe_4Sb_{12}$

Juan Zhou, Qing Jie, Lijun Wu, Ivo Dimitrov, Qiang Li[a]

*Condensed Matter Physics and Materials Science Department, Brookhaven National Laboratory, Upton, NY 11973, USA*

Xun Shi

*CAS Key Laboratory of Materials for Energy Conversion, Shanghai Institute of Ceramics, Chinese Academy of Sciences, 1295 Dingxi Road, Shanghai 200050, China*

## ABSTRACT

We studied nanoprecipitates and defects in p-type filled skutterudite $CeFe_4Sb_{12}$ prepared by non-equilibrium melt-spinning plus spark plasma sintering method using transmission electron microscopy. Nanoprecipitates with mostly spherical shapes and different sizes (from several nm to several tens of nm) have been observed. The most typically observed nanoprecipitates are shown to be Sb-rich. Superlattices with a periodicity of about 3.576 nm were induced by the ordering of excessive Sb atoms along the *c* direction. These nanoprecipitates usually share coherent interfaces with the surrounding matrix and induce anisotropic and strong strain fields in the surrounding matrix. Nanoprecipitates with compositions close to $CeSb_2$ are much larger in size (~ 30 nm) and have orthorhombic structures. Various defects were typically observed on the interfaces between these nanoprecipitates and the matrix. The strain fields induced by these nanoprecipitates are less distinct, possibly because part of the strains has been released by the formation of defects.

[a] Address all correspondence to this author.
E-mail: qiangli@bnl.gov

# I. INTRODUCTION

The efficiency of a thermoelectric (TE) material is measured by the dimensionless figure of merit $ZT = S^2\sigma T/\kappa$, where $S$, $\sigma$, $T$ and $\kappa$ stand for the Seebeck coefficient, electrical conductivity, absolute temperature and thermal conductivity, respectively.[1] The total thermal conductivity, $\kappa$, includes two parts: the electrical contribution, $\kappa_e$ and the lattice contribution, $\kappa_L$, the latter of which can be tuned almost independently of other parameters. Filled skutterudites are typical materials for the phonon glass electron crystal (PGEC) paradigm,[2] and they have been shown to exhibit much lower $\kappa_L$ than their parent compounds. Filler atoms, such as rare earth,[3-4] alkali,[5] or alkaline earth[6] elements have been utilized to fill the cages of the open host skutterudite structures in order to cause localized rattling, superimposed on the collective motions of lattice plane waves, leading to the scattering of heat-carrying acoustic phonons. In addition to filling parent compounds, nanostructuring has been shown to play an important role in reducing the thermal conductivities of a variety of thermoelectric materials.[7-9]

Non-equilibrium synthesis techniques such as rapid solidification and fast sintering are known for producing materials with nanostructures. Recently, a non-equilibrium synthesis method combining melt spinning with subsequent spark plasma sintering (SPS) or hot press has been successfully employed in the preparation of p-type filled skutterudites[10-11] and several other thermoelectric materials. Non-equilibrium synthesized n-type $Yb_{0.2}Co_4Sb_{12+y}$ samples with excessive Sb showed dispersion of nanoscale secondary phase on the grain boundaries.[12] The nanostructures were believed to have reduced $\kappa_L$ and improved the electrical properties at the same time.[12]

Non-equilibrium synthesis provides a fast and economical way of processing bulk thermoelectric materials with superior properties. Currently, detailed structural information such as size, shape, crystal structure and distribution of nanoprecipitates in the non-equilibrium synthesized materials, as well as defects and strain field induced by nanoprecipitates, which substantially influence thermoelectric properties, is still lacking. In this paper, we report our combined high resolution transmission electron microscopy (HRTEM) imaging with chemical composition, lattice mismatch and strain field analysis of nanoprecipitates in the non-equilibrium synthesized $CeFe_4Sb_{12}$ samples.

## II. EXPERIMENTAL PROCEDURE

### A. Non-equilibrium synthesis of $CeFe_4Sb_{12}$ polycrystals

High purity element pieces of Ce (99.8%), Fe (99.98%) and Sb (99.9999%) at stoichiometric amounts were loaded into boron nitride (BN) tubes, which were then sealed under Ar pressure. The raw materials were melted at about 1450 °C for 30 seconds in an induction furnace, and then cooled down to room temperature in 30 minutes. The obtained ingot was placed into a quartz tube with a 0.5 mm diameter nozzle, melted and injected under an Ar pressure of 0.067 MPa Ar onto a copper wheel rotating at a linear speed of 30 m/s. The melt-spun $CeFe_4Sb_{12}$ ribbons were pressed into a pellet and densified in the spark plasma sintering (SPS) system in vacuum under 50 MPa at 600 °C for 5 minutes.

### B. High resolution transmission electron microscopy (HRTEM)

Samples for transmission electron microscopy (TEM) observation were prepared by a traditional dimpling method. After mechanical polishing and dimpling to a thickness of about 20 μm, the samples were thinned to electron transparent using an ion mill system at low milling angles (6-12°). The specimen stage of the ion mill system was cooled with liquid nitrogen, which helped to avoid local specimen overheating during the milling process to reduce damage to the sample. All TEM samples were ion milled below -90 °C. HRTEM was performed on a JEOL JEM 2100F transmission electron microscope equipped with a Schottky field-emission gun and two exchangeable objective-lens pole-pieces. The ultra high resolution pole piece has a 0.19 nm point-to-point resolution and a ± 20° sample tilt, and the high resolution pole piece has a 0.23 nm point-to-point resolution and a ± 40° sample tilt. It is also equipped with an Oxford energy dispersive X-ray spectrometer (EDS) for chemical analysis.

The local distribution of the strain field around nanoprecipitates was retrieved using the geometric phase analysis (GPA) software. GPA is capable of generating quantitative deformation and strain maps from standard HRTEM images based on the geometric phase algorithms originally developed by Martin Hytch.[13] The first step is to calculate the Fourier transform of the HRTEM image.[7] Phase image is obtained by centering a small aperture on a strong reflection spot $g$, followed by an inverse Fourier transform. The phase component $P'_g(r)$ of the phase image is related to the lattice displacement $u(r)$ by

$$P_g(r) = P'_g(r) - 2\pi g \cdot r = -2\pi g \cdot u(r)$$

where $g$ is the reciprocal lattice vector from the undistorted reference lattice. The two-dimensional displacement field can be derived by applying the method to two non-collinear Fourier components, e.g. $g_1$ and $g_2$.

$$P_{g1}(r) = -2\pi g_1 \cdot u(r) = -2\pi[g_{1x}u_x(r) + g_{1y}u_y(r)]$$

$$P_{g2}(r) = -2\pi g_2 \cdot u(r) = -2\pi[g_{2x}u_x(r) + g_{2y}u_y(r)]$$

where $g_{1x}$, $g_{1y}$, $g_{2x}$ and $g_{2y}$ are the $x$ and $y$ components of the $g_1$ and $g_2$ vectors,

respectively. $u_x(r)$ and $u_y(r)$ are the $x$ and $y$ components of the displacement field $u(r)$ at the position $r = (x, y)$ in the image. The strain field is then calculated by

$$\varepsilon_{xx} = \partial u_x/\partial x, \; \varepsilon_{yy} = \partial u_y/\partial y$$

## III. RESULTS AND DISCUSSION

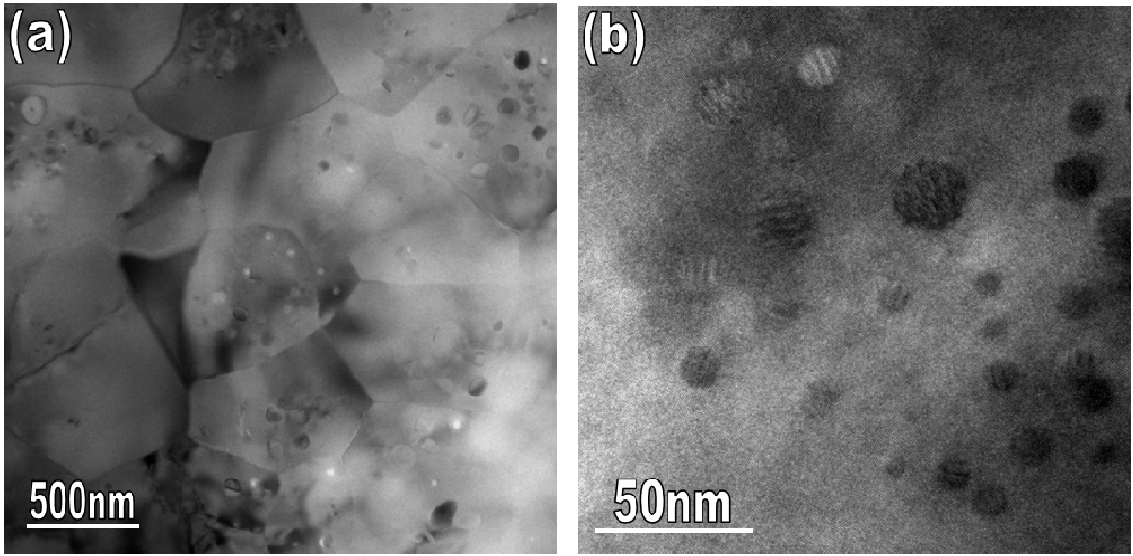

FIG. 1. TEM images at low (a) and high (b) magnifications showing nanometer sized grains of the non-equilibrium synthesized $CeFe_4Sb_{12}$ samples containing abundant nanoprecipitates.

The grains of the experimental samples have been reported to vary from several hundred nanometers to around 1 μm.[14] Individual grains containing abundant nanoprecipitates were readily observed in the samples, as shown in Fig. 1(a). The TEM image in Fig. 1(b) shows the kind of nanoprecipitates which are most typically observed. They have spherical shapes with coarse fringes. Further HRTEM imaging shows that they are typically coherently embedded in the matrix. The size of this kind of nanoprecipitates varies from a few nm to less than 20 nm.

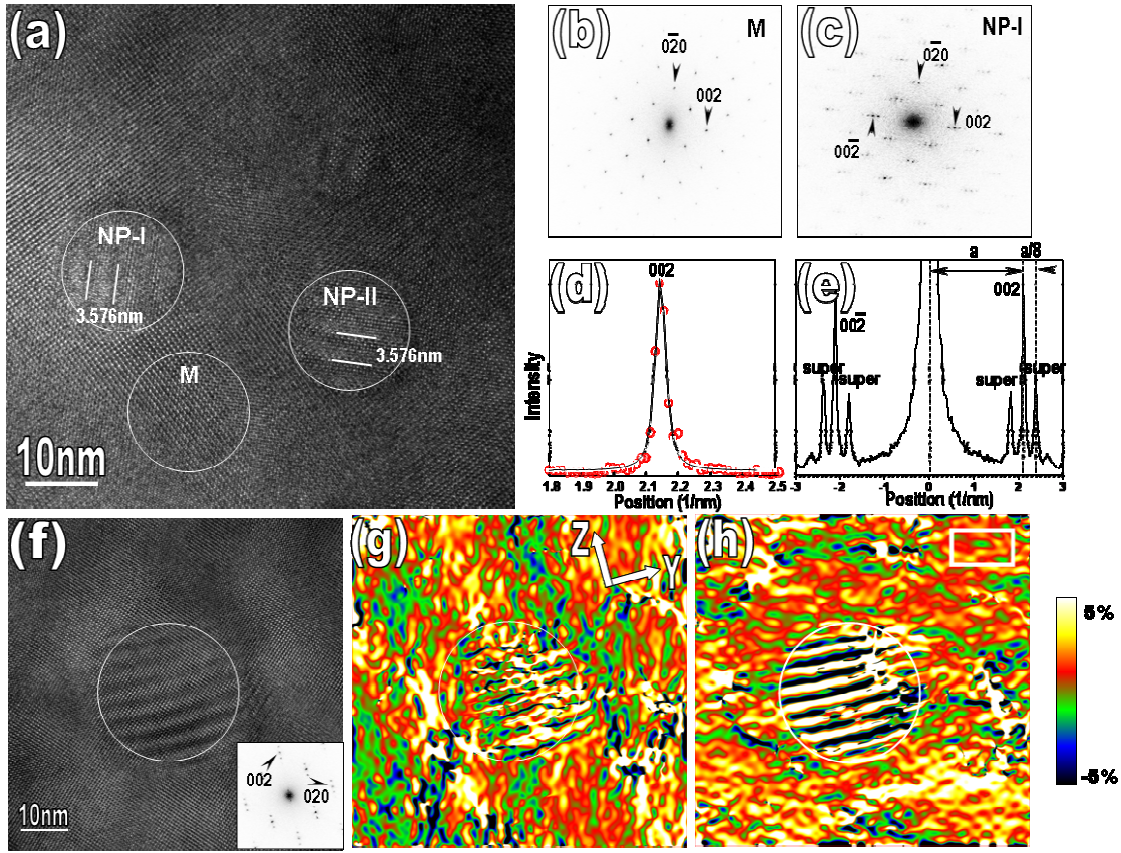

FIG. 2. (a) An HRTEM image of the studied $CeFe_4Sb_{12}$ sample viewed along the [100] zone axis, showing two nanoprecipitates coherently embedded in the matrix, designated as NP-I and NP-II. The fast Fourier transform (FFT) diffractograms of the circled matrix area M and nanoprecipitate area NP-I in (a) are shown in (b) and (c), respectively. In (c), the presence of superlattice reflections spots is evident. (d) Combined Gaussian and Lorentzian fitting (black solid line) to the 002 peak of the matrix (red open circles). (e) Line scan of intensity profiles from 00-2 peak to 002 peak of the diffractogram of NP-I in (c). (f) An HRTEM image with a nanoprecipitate embedded. FFT diffractogram of the circled nanoprecipitate region is shown in the inset. (g) - (h) Strain maps of $\varepsilon_{yy}$ and $\varepsilon_{zz}$, which were calculated from the HRTEM image shown in (f). The white rectangle in (h) is the reference area. The maps are shown in color for clarity. The circles in the figures outline the nanoprecipitate.

Fig. 2(a) is an HRTEM image viewed along the [100] direction, showing two such kind of nanoprecipitates coherently embedded in the matrix of a single grain. The two nanoprecipitates are about 15 nm in diameter and show fringes with the spacing being about 3.576 nm. No obvious lattice distortion was observed on the interfaces between the nanoprecipitates and the surrounding matrix. Here, we denote the two nanoprecipitates as NP-I and NP-II (See Fig. 2(a)). The fast Fourier transform (FFT) diffractograms from the circled matrix area M and the nanoprecipitate area NP-I are shown in Figs. 2(b) and 2(c), respectively. The matrix is confirmed to be the body-centered cubic (BCC) filled skutterudite structure. EDS measurements show that the concentration of Sb in the NP-I and NP-II areas is significantly higher than that of the neighboring matrix. NP-I and NP-II areas are determined to be an incommensurate superstructure, in which excessive Sb atoms order along the matrix *c* direction and induce superlattice reflection spots (Fig. 2(c)). Fig. 2(d) shows the line scan of the 002 peak intensity profile from the matrix (red open circles). Accurate lattice parameter was obtained by fitting the profile using a combination of a Gaussian and a Lorentzian function. The black solid line in Fig. 2(d) is the fitting of the 002 matrix peak, from which the lattice parameter of the matrix is determined to be 0.918 ± 0.003 nm. The line scan of the intensity profile from 00-2 peak to 002 peak of the FFT diffractogram (Fig. 2(c)) of NP-I area is shown in Fig. 2(e). The spacing between the main spots and the superlattice spots is measured to be 0.28 $nm^{-1}$, indicating that the periodicity of the superlattice in NP-I area is about 4 times larger than that of the matrix. Since the matrix has the cubic structure, whose *a*, *b*, and *c* axes are equivalent, the ordering of Sb atoms can be along *a*, or *b* direction of the matrix as well. As marked in Fig. 2(a), the periodicity of superlattice in NP-II area is same to that of NP-I area as 3.576 nm, except that the ordering of superlattice in NP-II is along the *b* direction.

Strain field distribution retrieved by the GPA software is a relative strain map with respect to the reference area. Fig. 2(f) shows an HRTEM image of a nanoprecipitate, in which the FFT diffractogram from the circled nanoprecipitate area is shown in the inset.

Figs. 2(g) and 2(h) show the relative strain maps, $\varepsilon_{yy}$ and $\varepsilon_{zz}$, retrieved from the HRTEM image in Fig. 2(f) with $g_1$ = 020 and $g_2$ = 002, and $y$ and $z$-axes pointing to the [010] and [001] directions. The reference area is marked by a white rectangle in Fig. 2(h). The $\varepsilon_{yy}$ map exhibits both positive and negative strains, and the $\varepsilon_{zz}$ map shows strong positive strains in the matrix. Although the nanoprecipitate has a symmetric (circular) shape, $\varepsilon_{yy}$ and $\varepsilon_{zz}$ strains around nanoprecipitate are anisotropic. There are small regions near the nanoprecipitate showing positive values of $\varepsilon_{xx}$ and $\varepsilon_{zz}$ is larger on the bottom side than the top side. The anisotropic distribution of strain field around nanoprecipitates with symmetric shapes has also been observed in AgPb$_m$SbTe$_{m+2}$ (LAST-18) single crystals.[7] One possible explanation lies in that strain fields are not only caused by the misfit between nanoprecipitates and the matrix, but also by the local fluctuation in composition. The formation of Sb-rich nanoprecipitates consumes Sb, which naturally induces local compositional fluctuation in the surrounding matrix. It is also possible that the formation of the Sb-rich superlattices may be caused by the Ce-filling fraction fluctuation since a very quick sintering was used in order to inhibit grain growth. The alternate contrast inside the nanoprecipitate areas was formed by the modulated superlattice structures.

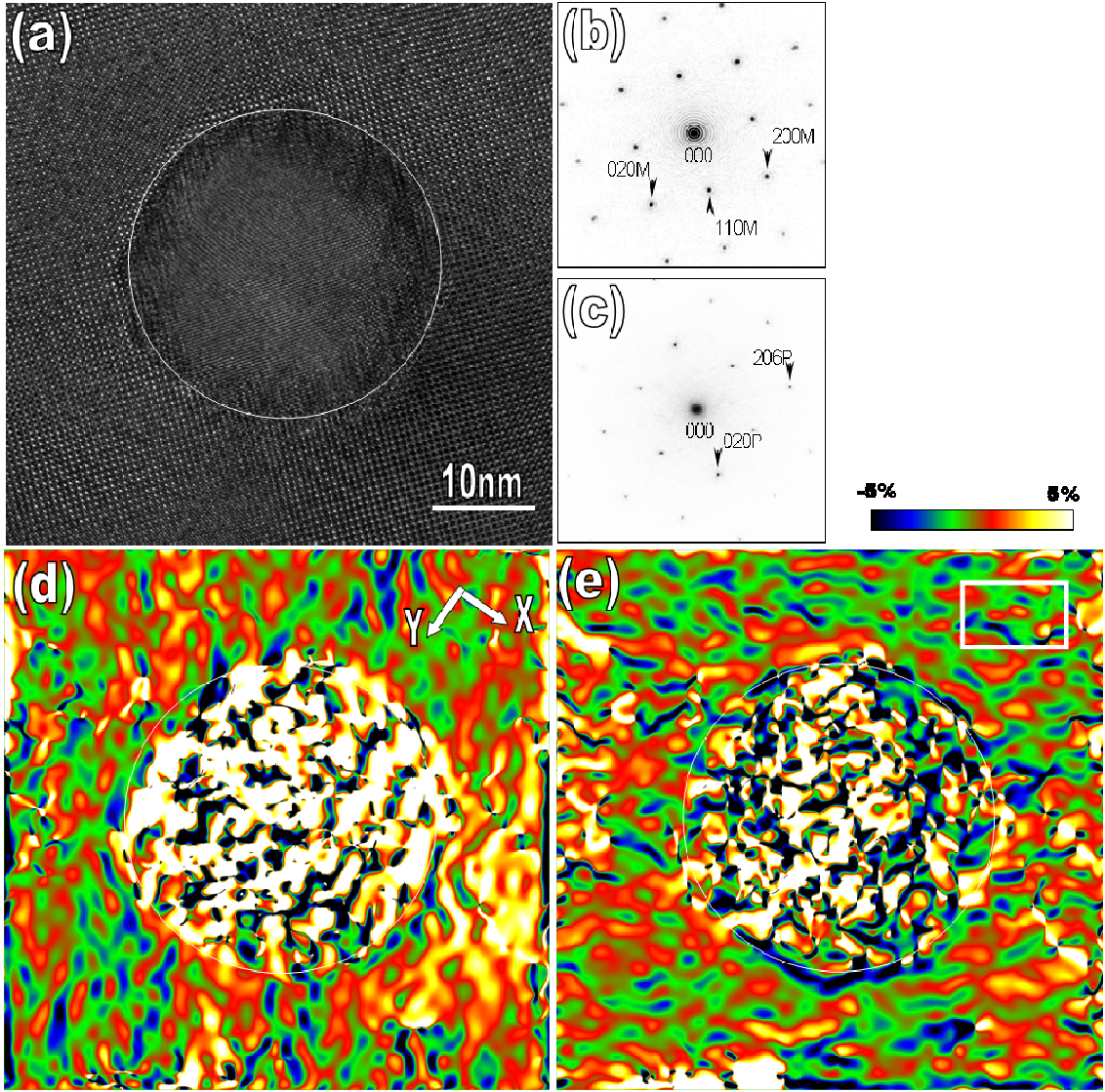

FIG. 3. (a) HRTEM image showing a CeSb$_2$ nanoprecipitate embedded in the matrix of the non-equilibrium synthesized CeFe$_4$Sb$_{12}$ sample. Corresponding FFT diffractograms from the circled areas in the matrix and the nanoprecipitate in (a) are shown in (b) and (c), respectively. Strain-maps of (d) $\varepsilon_{xx}$ and (e) $\varepsilon_{yy}$ are calculated from the HRTEM image in (a). The white rectangle in (e) is the reference area. The circles in the figures outline the nanoprecipitates.

Fig. 3(a) shows the HRTEM image viewed along the [001] matrix direction, showing another typical kind of nanoprecipitates in the studied CeFe$_4$Sb$_{12}$ samples. It has a spherical shape and is about 30 nm in diameter. EDS analysis shows that the composition of the nanoprecipitate is close to CeSb$_2$. The FFT diffractograms from the circled matrix and nanoprecipitate areas are shown in Figs. 3(b) and 3(c), respectively. The nanoprecipitate was indexed to be a $CeSb_2$ orthorhombic structure with lattice parameters of $a = 0.6295(6)$ nm, $b = 0.6124(6)$ nm, and $c = 1.821(2)$ nm. The orientation relationship between the nanoprecipitate and the matrix was determined to be $(010)_{CeSb2}//(110)_{CeFe4Sb12}$ and $(206)_{CeSb2}//(1-10)_{CeFe4Sb12}$, respectively. A possible $CeSb_2$ phase was not detected from the X-ray diffraction pattern,[14] indicating that only trace amounts of it exist in the sample. In Fig. 3(a), lattice distortions and misfit dislocations were frequently observed surrounding the nanoprecipitate, which are possibly formed to compensate the lattice mismatch between the nanoprecipitate and the neighboring matrix.

Figs. 3(d) and 3(e) show the relative strain maps $\varepsilon_{xx}$ and $\varepsilon_{yy}$, retrieved from the HRTEM image in Fig. 3(a) with $g_1 = 200$, $g_2 = 020$, and $x$, $y$ axes pointing to the [100] and [010] directions. The reference area is marked by a white rectangle in Fig. 3(e). The $\varepsilon_{xx}$ map shows positive strain in both the matrix and the nanoprecipitate, which is smaller on the top left than on the bottom right. The $\varepsilon_{yy}$ map shows positive strain in the nanoprecipitate and almost no strain in the matrix. The strains maps are also anisotropic and $\varepsilon_{yy}$ shows much less strain than $\varepsilon_{xx}$. Compared to the Sb-rich nanoprecipitates which share coherent interfaces with the matrix, the $CeSb_2$ nanoprecipitates have defects on the interfaces and produce less distinct strains on the matrix. The possible reason may be that part of the strains caused by the $CeSb_2$ nanoprecipitate has been released through the formation of dislocations and other defects on the interfaces.

## IV. CONCLUSIONS

The nanoprecipitates in the non-equilibrium synthesized $CeFe_4Sb_{12}$ have been analyzed in detail using quantitative HRTEM. The Sb-rich nanoprecipitates usually share coherent interfaces with the matrix. The large $CeSb_2$ nanoprecipitates were observed with lattice distortions and misfit dislocations on the interfaces with the matrix. The strain fields in the matrix around these spherical nanoprecipitates are anisotropic. Coherent interfaces between nanoprecipitates and the matrix seem to induce more distinct strain fields than interfaces with defects. The formation of defects on the interface may have released part of the strains. The high density of additional interfaces and strain fields produced by the nanometer sized grains, nanoprecipitates, and defects are believed to enhance phonon scattering and result in a significant reduction in $\kappa_L$. Currently, *ab initio* density functional theory and lattice dynamical models are being aimed at modeling the strain formation around the nanoprecipitates which were observed in the HRTEM imaging of our sample and future inelastic neutron scattering measurements are planned to observe potential effects of nanostructuring on the phonon density of states in a number of filled skutterudites.

## ACKNOWLEDGEMENTS

The work at Brookhaven National Laboratory was primarily supported by the Office of Science, U.S. Department of Energy, under Contract No. DE-AC02-98CH10886. We thank Dr. Jihui Yang of Materials and Processes Laboratory, General Motors R&D Center for generous help with sample preparation. This research was also partly supported by the CRADA between Brookhaven National Laboratory (Q. Li) and General

Motors Corporation (J. H. Yang). We thank the Center for Functional Nanomaterials, Brookhaven National Laboratory for generous support in using its facilities.**REFERENCES**

1. H. J. Goldsmid, *Thermoelectric Refrigeration*. 1964, New York: Plenum Press.
2. G. A. Slack, ed. *CRC Handbook of Thermoelectrics*. ed. D.M. Rowe. 1995, CRC Press: Boca Raton.
3. D. T. Morelli, and G.P. Meisner: Low temperature properties of the filled skutterudite CeFe$_4$Sb$_{12}$. *J. Appl. Phys.* **77**, 3777 (1995)
4. G. S. Nolas, G. A. Slack, D. T. Morelli, T. M. Tritt, and A. C. Ehrlich: The effect of rare-earth filling on the lattice thermal conductivity of skutterudites. *J. Appl. Phys.* **79**, 4002 (1996)
5. Y. Z. Pei, L. D. Chen, W. Zhang, X. Shi, S. Q. Bai, X. Y. Zhao, Z. G. Mei, and X. Y. Li: Synthesis and thermoelectric properties of K$_y$Co$_4$Sb$_{12}$. *Appl. Phys. Lett.* **89**, 221107 (2006)
6. L. D. Chen, T. Kawahara, X. F. Tang, a.T.H. T. Goto, IJ. S. Dyck, W. Chen, and C. Uher: Anomalous barium filling fraction and n-type thermoelectric performance of Ba$_y$Co$_4$Sb$_{12}$. *J. Appl. Phys.* **90**, 1864 (2001)
7. L. J. Wu, J. C. Zheng, J. Zhou, Q. Li, J. H. Yang, and Y.M. Zhu: Nanostructures and Defects in Thermoelectric AgPb$_{18}$SbTe$_{20}$ Investigated by Quantitative High Resolution Transmission Electron Microscopy. *J. Appl. Phys.* **105**, 094317 (2009)
8. K. F. Hsu, S. Loo, F. Guo, W. Chen, J. S. Dyck, C. Uher, T. Hogan, E. K. Polychroniadis, and M. G. Kanatzidis: Cubic AgPb$_m$SbTe$_{2+m}$: Bulk thermoelectric materials with high figure of merit. *Science*. **303**, 818 (2004)
9. B. Poudel, Q. Hao, Y. Ma, Y. C. Lan, A. Minnich, B. Yu, X. Yan, D. Z. Wang, A. Muto, D. Vashaee, X. Y. Chen, J. M. Liu, M. S. Dresselhaus, G. Chen, and Z. F. Ren: High-Thermoelectric Performance of Nanostructured Bismuth Antimony Telluride Bulk Alloys. *Science*. **320**, 634 (2008)
10. Q. Li, Z. W. Lin, and J. Zhou: Thermoelectric Materials with Potential High Power Factors for Electricity Generation. *J Electron Mater*. **38**, 1268 (2009)
11. Q. Jie, J. Zhou, I. K. Dimitrov, C. P. Li, C. Uher, H. Wang, W. Porter, and Q. Li: Thermoelectric Properties of Non-equilibrium Synthesized Ce$_{0.9}$Fe$_3$CoSb$_{12}$ Filled Skutterudites. *Mater. Res. Soc. Symp. Proc.* **1267**, DD03-03 (2010)
12. H. Li, X. F. Tang, X. L. Su, and Q.J. Zhang: Preparation and thermoelectric properties of high-performance Sb additional Yb$_{0.2}$CoSb$_{12+y}$ bulk materials with nanostructures. *Appl. Phys. Lett.* **92**, 202114 (1-3) (2008)
13. M. J. Hÿtch, E. Snoeck, and R. Kilaas: Quantitative measurement of displacement and strain fields from HREM micrographs. *Ultramicroscopy*. **74**, 131 (1998)
14. J. Zhou, Q. Jie, and Q. Li: Microstructure of non-equilibrium synthesized p-type Filled Skutterudite CeFe$_4$Sb$_{12}$. *Mater. Res. Soc. Symp. Proc.* **1267**, DD05-25 (2010)
12